\documentclass[12pt]{amsart}
\usepackage{graphicx}
\usepackage[margin = 1in]{geometry}
\begin{document}
\title{New Time Series Models for Corporate Bond Log Yields}
\author{Jihyun Park, Andrey Sarantsev}

\begin{abstract}
We propose a class of simple time series models for rates and spreads of portfolios of corporate bonds classified by ratings provided by Bank of America. We evaluate these models based on statistical analysis of innovations: Whether they are independent identically distributed. Our tests are unusually rigorous, compared with standard practice, Our novelty is taking logarithms of rates or spreads instead of rates or spreads themselves. We find out that the best option is to take logarithms of spreads of logarithms of rates. Sometimes, dividing these innovations by the volatility index for stocks makes them pass our statistical tests. It is remarkable that stock volatility can also serve as bond volatility. 
\end{abstract}

\maketitle

\thispagestyle{empty}

\section{Introduction} 

\subsection{Risk factors: existing literature} There is a large literature on studying risk factors for corporate bond yields and returns, but it is focused mostly on adding and comparing individual factors, not building comprehensive time series models with analysis of residuals. See for example recent articles \cite{Avramov, Kurtosis, Kagraoka}, where they study risk factors for corporate bond spreads. Of course, we need to mention the classic article \cite{FF1993}, usually cited with regard to the size and value effect for stock portfolios, but they also study two risk factors for corporate bonds. Also, the article \cite{GUO} connects corporate bond rates with their total returns. A remarkable recent work was done in \cite{Pricing}, with a century-long database for corporate bonds. We do not intend to provide a comprehensive review of literature here, since this is but a short note, intended only to introduce our new idea. 

\subsection{General discussion of time series models} These include classic random walk and autoregressions, as well as more complicated stochastic volatility or GARCH-type equations. There is a large literature on fitting and studying such models, see for example the monograph \cite{garch}. A particular question of interest is studying their stationarity: For example, unit root tests distinguishing between a random walk and an autoregression, or finding an optimal order of an autoregression. See the classic textbook \cite{Brockwell-Davis}. The models are evaluated based on their various statistical characteristics: information criteria (AIC, BIC), $R^2$, normality test for innovations, and log-likelihood. 

But it is hard to find comprehensive analysis of innovations for white noise. This is unfortunate, because such testing is important. Time series models assume the innovations in equations are independent identically distributed (IID). Sometimes, there is no testing at all. Or only the Durbin-Watson test is applied to innovations: This tests the autocorrelation function for lag 1 only. The most compehensive approach usually taken is plotting or analyzing the autocorrelation function for these innovations. 

\subsection{Motivation for testing absolute values of innovations} But even when autocorrelations for all or most lags are not significantly different from zero, this is not enough to conclude that said innovations are indeed IID. Indeed, it is common to model stock market return series having zero autocorrelations but stochastic volatility. This implies that squares of returns (or absolute values of returns) do have significant autocorrelations. See, for example, Heston or GARCH models. See the popular textbook \cite{fan}.

The intuition behind this: If today's returns are significantly correlated with yesterday's returns, then this can be easily exploited by traders, and so it will be arbitraged away. But periods of crisis alternate with periods of calm in financial markets. During the former, volatility is high, but during the latter, it is low. Therefore, we can conclude: If yesterday's returns had high magnitude (absolute value), then it is very likely that today's returns also had high magnitude.

Thus we need to apply empirical ACF not just for innovations $Z_t$, but to their absolute values $|Z_t|$ as well. Only such rigorous testing can distinguish true IID data from non-IID models with zero autocorrelation. Only this can filter out stochastic volatility models such as Heston or GARCH. See more in \cite{IID}. Alternative ways of testing are provided in \cite{brock, mcleod}.

\subsection{Stochastic volatility for stock returns} These models can be fit in various ways: (a) The volatility is not observed and is inferred from observed returns as a hidden variable. This is how classic GARCH models are fit, see \cite{garch}. (b) The volatility is evaluated separately. The latter case has two sub-cases: (i) the standard deviation of daily returns (realized volatility); (ii) the solution of the Black-Scholes equation for option prices with volatility as the unknown variable (implied volatility). For the (b) (ii) case, we need a liquid option market for a given stock or an index. Such market is particularly rich for S\&P 500, and is used for constructing the celebrated Volatility Index (VIX), see \cite{ang}. The daily data for VIX is available since the start of 1990. As an example, one can point to a recent preprint \cite{LogHeston}, where they discuss dividing monthly stock returns by VIX to make them closer to IID Gaussian. 

\subsection{Stochastic volatility for corporate bond rates} The same intuition can be applied to bond markets. Whether we model the bond rates as autoregression (a mean-reverting, stationary model), or as a random walk (a non-stationary model), we could use either (a) or (b) approach. For corporate bonds, there also exist options, and we can infer implied volatility from these. This index is known as MOVE. We can view this as a cousin of VIX in the bond market. But this is available only from 2002, so we shall use VIX instead. Also, we downloaded VIX data from the (very well-organized) FRED web site, but struggled to find MOVE historical data in a convenient format. 

The decision to use VIX for bond instead of stock markets is controversial, and we defend it by noting that bond and stock markets are closely related and inter-dependent. This is not surprising; after all, bond and stock markets compete for many of the same investors. In particular, a change in bond rates immediately impacts stock prices. We understand some readers might not agree with our approach, and we leave this for future research. This short manuscript is only exploratory; its ideas must be tested further. 

\begin{figure}[t]
\includegraphics[width = 10cm]{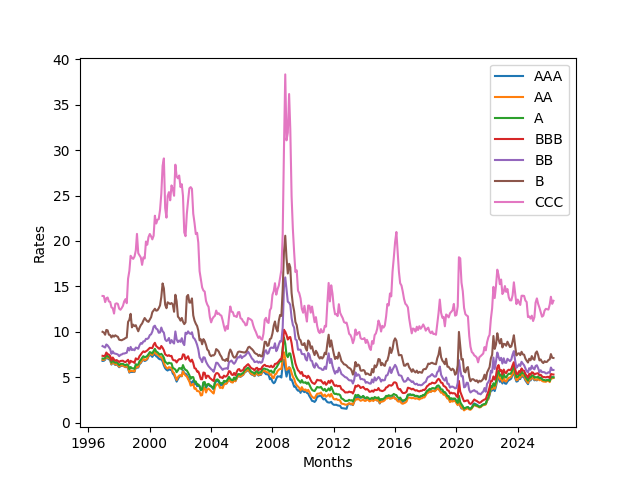}
\caption{Bond Rates}
\end{figure}

\subsection{Motivation for taking logarithms} Applying logarithms is standard practice in stock market analysis for wealth and price processes. Indeed, one justification is that wealth and prices are always positive (except the unfortunate case of bankruptcy), so one takes logarithms to make the range of possible values $(-\infty, +\infty)$ instead of $(0, +\infty)$. However, for the bond markets, such logarithmic transform is much less common in the literature. For example, principal component analysis can be applied to a series of bond rates. The classic example is the yield curve, which has Treasury rates with various maturities, from short to long. The first principal component is the level, the second one is the slope, etc. See \cite{ang}. However, to the best of our knowledge, no one has applied this method to logarithms of rates instead of rates themselves. This is even more surprising given that, in this case, it would make perfect sense: All bond rates are always positive. The PCA is naturally suited (as a linear transformation) for data which has range $(-\infty, +\infty)$ rather than $(0, +\infty)$. 

What is more, credit spreads (difference between a lower-rated bond and a higher-rated bond with the same maturity) are positive as well. Indeed, high credit risk of the lower-rated bond must be compensated by its higher rate, \cite{ang}. Thus it makes sense to take the logarithm of said spread as well. Going even further, we can take spread of log rates, or log spread of log rates. All these options are listed above. 

\section{Methodology of Testing for IID} For the time series $Z_1, \ldots, Z_T$, which we want to test for IID, the test statistic is 
$$
\hat{Q}_T(k) = T(T+2)\sum\limits_{j=1}^k\frac{\hat{\rho}^2_T(j)}{T - j}
$$
where we define the empirical autocorrelation function (ACF) as
$$
\hat{\rho}_T(j) = \frac1T\sum\limits_{i=1}^{T-j}(Z_i - \overline{Z})(Z_{i+j} - \overline{Z}),
$$
where, in turn, $\overline{Z} := (Z_1 + \ldots + Z_T)/T$. It is known that, as $T \to \infty$, the limiting distribution is $\chi^2_k$ for $\hat{Q}_T(k)$ for each $k$. The division by $T - j$ and multiplication by $T+2$ was introduced for faster convergence. This known limiting $\chi^2$ distribution allows us to compute the $p$-values. This is all quite standard, see \cite{Brockwell-Davis}. 

{\it Our rule:} The test itself gives us four $p$-values: Ljung-Box test for zero autocorrelation function for $k = 5$ lags and $k = 10$ lags for each of the two series $Z_t$ and $|Z_t|$. The series passed the test if and only if all four $p$-values are greater than $1\%$ (that is, we fail to reject the null hypothesis that the series $Z_t$ or $|Z_t|$ has zero autocorrelation function). We make this threshold lower than the usual 5\% because of multiple simultaneous hypotheses testing. We wish to cast the net wide and record even those cases with low but non-negligible $p$-values. 

\section{Monthly Data 1996--2025}

We take end-of-month Bank of America Intercontinental Exchange (BofA ICE) rates for ratings: AAA, AA, A, BBB, BB, B, CCC, for December 1996 -- May 2026 time period. We denote this rate for the rating $r$ as $R_r(t)$ for the month $t$. Also, the monthly average daily VIX for the same time period is denoted by $V(t)$. The data is taken from Federal Reserve Economic Data (FRED) web site: 

\begin{itemize}
\item AAA rates: ICE BofA AAA US Corporate Index Effective Yield \newline series code BAMLC0A1CAAAEY
\item AA rates: ICE BofA AA US Corporate Index Effective Yield \newline series code BAMLC0A2CAAEY
\item A rates: ICE BofA Single-A US Corporate Index Effective Yield \newline series code BAMLC0A3CAEY
\item BBB rates: ICE BofA BBB US Corporate Index Effective Yield \newline series code BAMLC0A4CBBBEY
\item BB rates: ICE BofA BB US High Yield Index Effective Yield \newline series code BAMLH0A1HYBBEY
\item B rates: ICE BofA Single-B US High Yield Index Effective Yield \newline series code BAMLH0A2HYBEY
\item CCC rates: ICE BofA CCC \& Lower US High Yield Index Effective Yield \newline series code BAMLH0A3HYCEY
\item VIX: CBOE Volatility Index: VIX (series code VIXCLS)
\end{itemize}

Currently, the data for each series of rates are available from FRED only for the last three years. But we have downloaded them back when they provided a full daily history starting from the last trading day of 1996. Since then, we have updated the data. The data for VIX  is available daily since 1990 without any restrictions. The data files are in \texttt{GitHub} repository \texttt{asarantsev/log-spreads} 

\section{Time Series} For any time series $X_0, X_1, \ldots, X_T$, we model it in two ways:

\begin{enumerate}
\item As a {\it random walk} (RW): Equivalently, $Z_t = X_t - X_{t-1}$ are IID; 
\item As an {\it autoregression of order 1} AR(1): $X_t = a + bX_{t-1} + Z_t$, with $a$ and $b$ found using the standard ordinary least squares (OLS) procedure, and $Z_t$ are IID. 
\end{enumerate}

If $b = 1$, then the AR(1) reduces to RW. This brings the aforementioned unit root testing: Distinguish between the stationary autoregression and the non-stationary random walk. Equivalently, decide whether the estimated value of $b$ (usually slightly less than one but close to it) is significantly different from 1. This is a classic topic in time series analysis and econometrics (Dickey-Fuller and related tests, for example). But we shall bypass this in our short note. As mentioned before, this work intends to introduce new ideas. To investigate them comprehensively, we need much more efforts. 

\begin{figure}[t]
\includegraphics[width = 10cm]{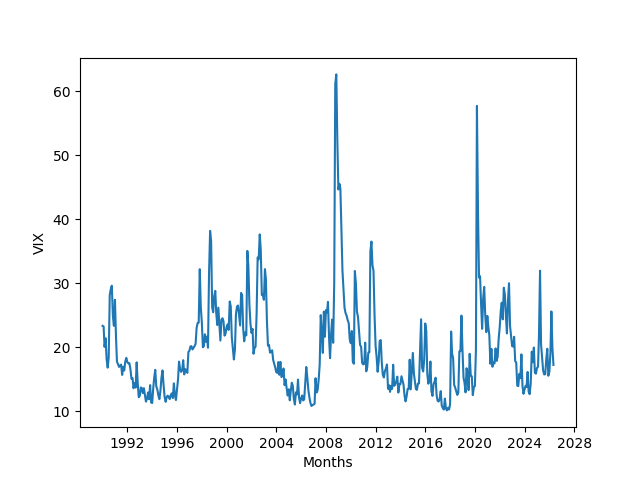}
\caption{The Volatility Index}
\end{figure}

\section{Rates and Spreads} For the series $X_t$, we take six possible cases. The first two are for bond rates for each rating $r$ out of AAA, AA, A, BBB, BB, B, CCC: 

\begin{itemize}
\item $R_r(t)$ for each rating $r$
\item $\ln R_r(t)$ for each rating $r$
\end{itemize}

The remaining four are for bond spreads for each rating $r$ lower than AAA; that is, for $r$ which is AA, A, BBB, BB, B, CCC. 

\begin{itemize}
\item $S_{r, 0} = R_r(t) - R_{AAA}(t)$ 
\item $S_{r, 1} = \ln(R_r(t) - R_{AAA}(t))$ 
\item $S_{r, 2} = \ln R_r(t) - \ln R_{AAA}(t)$ 
\item $S_{r, 3} = \ln(\ln R_r(t) - \ln R_{AAA}(t))$ 
\end{itemize}

Furthermore, for each innovation series $Z_t$ we test this series itself, and after dividing by monthly average VIX $V(t)$ (so the new series is $W_t = Z_t/V(t)$).

\section{Statistical Analysis Results} We find out that no series of rates, either with logarithms or without, pass the test. As for the spreads, we have them pass the test in the following cases: 

\subsection{Logarithms of spreads} 

\begin{itemize}
\item $S_{1, BBB}$ as RW with VIX; 
\item $S_{1, B}$ as RW or AR(1) without VIX; 
\item $S_{1, CCC}$ as RW with VIX; 
\end{itemize}

\subsection{Logarithms of spreads of logarithms}

\begin{itemize}
\item $S_{3, BBB}$ as RW or AR(1) with VIX;
\item $S_{3, BB}$ as AR(1) without VIX;
\item $S_{3, B}$ as RW or AR(1) without VIX;
\item $S_{3, CCC}$ as RW or AR(1) with VIX;
\item $S_{3, CCC}$ as RW or AR(1) without VIX;
\end{itemize}

\section{Interpretation}

The observation that only after logarithmic transformation (at least once) we get positive results shows that, in fact, taking logarithms was a good idea. 

Sometimes RW passes, sometimes AR(1) passes, sometimes both models pass. We would not attach much importance to this. This might be simply a statistical illusion. 

Sometimes we need to normalize by dividing the innovations by VIX. Other times, we do not need this. Again, part of this might be a statistical illusion. However, for stock returns, division by VIX leads to returns passing statistical tests for IID. For bond rates, this effect might be less pronounced.

Lower-rated bonds pass these tests: BBB, BB, B, and especially CCC, but higher-rated bonds do not pass these. We could offer the following interpretation: Lower-rated bonds have rates which sharply rise during financial crises and stock market crashes. But this makes them behave like the stock market.

\section{Conclusions and Further Research} 

The evidence for the bond rates is clear: Simple models such as RW or AR(1) (which are Markov) fail these rigorous IID testing. This stands in constrast with much of existing literature, where lots of models are proposed but without rigorous testing for residuals or innovations. Taking logarithms or dividing by VIX does not help. 

However, bond credit spreads sometimes do pass these tests, but only if we take either log of the original spread or log of the spread of logs, and only for lower-rated bonds. This requires further investigations. 

One way might be to use MOVE bond volatility index, although it spans shorter time. Another direction is to work with foreign bonds, for example Canada or UK. This will help to prove or disprove that this holds across countries, which is important for reproducibility. Furthermore, one might use IID testing based on order statistic or rank correlation (Spearman). Finally, it is important to try to extend the time period. For some bond rates, data is available before 1996. The VIX is available since 1990, and its predecessor fo the S\&P 100 (which closely tracks the S\&P 500 in practice) is available since 1986. Alternatively, one could use realized bond volatility. 

Overall, this short note shows that taking logarithms of rates is a promising direction, but certainly more research is needed. 

\bibliographystyle{plain} 

\bibliography{new}

\begin{thebibliography}{10}

\bibitem{ang}
Andrew Ang.
\newblock {\em Asset Management: A Systematic Approach to Factor Investing}.
\newblock Oxford University Press, New York, NY, 2014.

\bibitem{Avramov}
Doron Avramov, Gergana Jostova, and Alexander Philipov.
\newblock Understanding changes in corporate credit spreads.
\newblock {\em Financial Analysts Journal}, 63(2):90--105, 2007.

\bibitem{brock}
William~A. Brock, W.~Davis Dechert, Jos{\'e}~A. Scheinkman, and Blake LeBaron.
\newblock A test for independence based on the correlation dimension.
\newblock {\em Econometric Reviews}, 15(3):197--235, 1996.

\bibitem{Brockwell-Davis}
Peter~J. Brockwell and Richard~A. Davis.
\newblock {\em Introduction to time series and forecasting}.
\newblock Springer Texts in Statistics. Springer-Verlag, New York, second
  edition, 2002.

\bibitem{Kurtosis}
Ephraim Clark and Selima Baccar.
\newblock Modelling credit spreads with time volatility, skewness, and
  kurtosis.
\newblock {\em Annals of Operations Research}, 262(2):431--461, 2018.

\bibitem{FF1993}
Eugene~F. Fama and Kenneth~R. French.
\newblock Common risk factors in the returns on stocks and bonds.
\newblock {\em Journal of Financial Economics}, 33(1):3--56, 2 1993.

\bibitem{fan}
Jianqing Fan and Qiwei Yao.
\newblock {\em The Elements of Financial Econometrics}.
\newblock Cambridge University Press, New York, 2017.

\bibitem{garch}
Christian Francq and Jean-Michel Zakoian.
\newblock {\em GARCH Models: Structure, Statistical Inference and Financial
  Applications}.
\newblock John Wiley \& Sons, 2 edition, 2019.

\bibitem{Pricing}
Mohammad Ghaderi, S{\'e}bastien Plante, Nikolai~L. Roussanov, and Sang~Byung
  Seo.
\newblock Pricing of corporate bonds: Evidence from a century-long
  cross-section, April 2025.
\newblock SSRN Working Paper No. 5218755.

\bibitem{GUO}
Xu~Guo, Hai Lin, Chunchi Wu, and Guofu Zhou.
\newblock Predictive information in corporate bond yields.
\newblock {\em Journal of Financial Markets}, 59:100687, 2022.

\bibitem{Kagraoka}
Yusho Kagraoka.
\newblock A time-varying common risk factor affecting corporate yield spreads.
\newblock {\em The European Journal of Finance}, 16(6):527--539, 2010.

\bibitem{mcleod}
A.~Ian McLeod and Wai~K. Li.
\newblock Diagnostic checking arma time series models using squared-residual
  autocorrelations.
\newblock {\em Journal of Time Series Analysis}, 4(4):269--273, 1983.

\bibitem{LogHeston}
Jihyun Park and Andrey Sarantsev.
\newblock Log heston model for monthly average vix.
\newblock {\em arXiv preprint}, arXiv:2410.22471, October 2024.

\bibitem{IID}
Andrey Sarantsev.
\newblock {I}{I}{D} time series testing.
\newblock {\em Theory of Stochastic Processes}, 27(43):41--52, 2023.

\end{thebibliography}

\end{document}